\documentclass{raa}

\usepackage{graphicx,times}             
\usepackage{natbib}
\input{epsf.sty}                        
\input{psfig.sty}  
\input{3parttable.sty}                     

\begin{document}

   \title{A Search for Metal-poor Stars Pre-enriched by Pair-instability Supernovae
\subtitle {I. A Pilot Study for Target Selection from Sloan Digital Sky Survey (SDSS)}
}

   \volnopage{Vol.0 (200x) No.0, 000--000}      
   \setcounter{page}{1}          

   \author{J. Ren
      \inst{1,2,3}
   \and N. Christlieb
      \inst{2}
   \and G. Zhao
      \inst{1,3}
   }
   \institute{The School of Space Science and Physics, Shandong University at Weihai, 
     Wenhua Xilu 180, 264209 Weihai, Shandong, China;\\
     \and
     Zentrum f\"ur Astronomie der Universit\"at Heidelberg, 
     Landessternwarte, K\"onigstuhl 12, D-69117 Heidelberg, Germany; 
  {\it N.Christlieb@lsw.uni-heidelberg.de}\\
     \and
     Key Lab of Optical Astronomy, National Astronomical Observatories, Chinese Academy of Sciences A20 Datun Road, Chaoyang,
     Beijing 100012, China. 
     {\it renjing@bao.ac.cn; gzhao@nao.cas.cn}
   }

   \date{Received~~2012 month day; accepted~~2012~~month day}

\abstract{ We report on a pilot study on identifying metal-poor stars
  pre-enriched by Pair-Instability Supernovae (PISNe). Very massive, first
  generation (Population III) stars ($140\,M_\mathrm{\sun} \leq M \leq
  260\,M_\mathrm{\sun}$) end their lives as PISNe, which have been predicted
  by theories, but no relics of PISNe have been observed yet.  Among the
  distinct characteristics of the yields of PISN, as predicted by theoretical
  calculations, are a strong odd-even effect, and a strong overabundance of Ca
  with respect to iron and the Solar ratio. We use the latter characteristic
  to identify metal-poor stars in the Galactic halo that have been
  pre-enriched by PISN, by comparing metallicites derived from strong,
  co-added Fe lines detected in low-resolution (i.e., $R=\lambda/\Delta\lambda
  \sim 2000$) spectra of the Sloan Digital Sky Survey (SDSS), with
  metallicities determined by the SDSS Stellar Parameters Pipeline (SSPP). The
  latter are based on the strength of the Ca~II~K line and assumptions on the
  Ca/Fe abundance ratio. Stars are selected as candidates if their metallicity
  derived from Fe lines is significantly lower than the SSPP
  metallicities. In a sample of 12,300 stars for which SDSS spectroscopy is
  available, we have identified 18 candidate stars. Higher resolution and
  signal-to-noise ratio spectra of these candidates are being obtained with
  the Very Large Telescope of the European Southern Observatory and the
  XSHOOTER spectrograph, to determine their abundance patterns, and to verify
  our selection method. We plan to apply our method to the data base of
  several million stellar spectra to be acquired with the LAMOST telescope in
  the next five years.
\keywords{stars: Population II, Population III -- stars: supernovae (PISNe) -- stars:
  abundance -- stars: chemically peculiar -- method: data analysis -- techniques: spectroscopic
 -- instrumentation: spectrographs}
}

   \authorrunning{Ren et al.}            
   \titlerunning{A Search for Metal-poor Stars Pre-enriched by
     Pair-instability Supernovae}  

   \maketitle

%
%
\section{Introduction}\label{sect:intro}

The first generation of stars (herafter Population III stars) in the Universe
formed several hundred million years after the Big Bang. If such stars have
survived until today, they can provide us via their abundance patterns with
information on the chemical composition of the early Universe.  Because the
formation of Pop.~III stars occurred in an environment absent of an efficient 
cooling agent (such as metals and/or dust, which are present at later times), 
they are believed to have been much more massive with characteristic masses 
of $>=$10 $M_{\sun}$, up to possibly several hundred solar masses \citep[see,
  e.g.,][]{bromm:99,bromm:09,abel:00,karlsson:11}, and can reflect the yields of the Big Bang
\citep[e.g.,][]{abel:02,bromm:02,nakamura:02}. This is also suggested by
simulations of the collapse of primordial molecular clouds
\citep{Ostriker:96}. However, some studies have suggested that lower mass stars 
could form within the gas cluster, due to the fragmentation of the turbulent 
primodial gas in the minihalos \citep[see e.g.,][]{clark:11a,clark:11b} or due to the 
protostellar feedback \citep{hosokawa:2011}.

Depending on the mass of the progenitor star, different physical mechanisms
lead to the explosion of Pop.~III stars at the end of their lives
\citep[see][]{bromm:03}. Their UV radiation contributes to the re-ionization
of the Universe, and they enrich their surroundings with metals
\citep[e.g.,][] {Gnedin:97, Ferrara:00,Madau:01,Mori:02,Yoshida:04}. Very
massive first stars ($140\,M_\mathrm{\sun} \leq M \leq 260\,M_\mathrm{\sun}$)
will explode as Pair-Instability Supernovae (PISNe), leading to complete
disruption of the progenitors \citep[e.g.,][]
{fryer:01,heger:02,lingenfelter:03}, enriching the interstellar medium (ISM,)
with heavy elements. Due to their short lives ($\leq 3\,My$), PISNe are likely
the first objects in the Universe that have enriched the interstellar medium
(ISM). In the progenitor stars of PISNe, after central helium burning, the
temperature and density in the core support pair-instability, and then a
partial collapse happens. The collapse proceeds to efficiently compress the
stars's core, until fast implosive oxygen and silicon burning take place due
to the overpressure, releasing enough energy to revert the collapse, and the
star is completely disrupted with no black hole or other remnants left behind
\citep{heger:02,fryer:01}. The recently observed object SN 2007bi
\citep{gal-yam:09} is hypothesized to have been a PISN.

The fraction of metal-poor stars that have formed directly from material that was
enriched by the yields of PISNe is predicted to be very small \citep{karlsson:08} and this
together with an observational bias may explain why no metal-poor star displaying
distinct PISNe signatures have been observed so far
\citep[e.g.,][]{christlieb:02,cayrel:04,cohen:04,barklem:05,frebel:05}. The
majority of Pop.~II stars with a dominant contribution from PISNe are
predicted to have Ca abundances in excess of $\mbox{[Ca/H]}=-2.6$, which implies
that a significant fraction of the PISN-dominant stars may have escaped 
detection \citep{karlsson:08}. In the past, wide-angle
spectroscopic surveys for metal-poor stars have used the Ca K line as a
metallicity indicator, relying on the fact that the vast majority of the
metal-poor stars follow a well-defined trend of [Ca/Fe] as a function of
{[Fe/H]} \citep[see][]{beers:99}. Due to the high [Ca/Fe] produced in
PISNe, the metallicity of Pop.~II stars pre-enriched by them would therefore
be overestimated by about a factor of 40 or more, and hence they would not be
selected as metal-poor candidates in these surveys.

The lower side of Fig.~4 of \citet{karlsson:08} shows a simulation of the
predicted distribution of stars for which $>90$\,\% of the total atmospheric
Ca abundance is synthesized in PISNe in the [Fe/Ca]-[Ca/H] plane, assuming the
number of PISNe occupied $10$\,\% of the primordial stellar population. The
observation data cover the entire metallicity regime of their model, well
reproducing the small scatter in observations \citep[see,
  e.g.,][]{cayrel:04,arnone:05,barklem:05}. Among the several hundred
 well-studied stars falling in the high number density regions in the
abundance ratio diagram, none shows the abundance signatures of PISNe. Note
that still below [Ca/Fe]$\sim-1$, there is a negligible fraction of PISN-enriched 
 stars. These are the stars that we are targeting with our selection method.

Non-rotating stars exploding as PISNe differ from core-collapse SNe in that
they exhibit a neutron excess in their interiors. Consequently, one of the
characteristic chemical signatures of PISNe is a strong odd-even effect, which
means particularly low abundance ratios of odd-$Z$ elements to even-$Z$
elements. Furthermore, the production of both rapid and slow neutron-capture
elements are absent due to the lack of excess neutrons in addition to shorter
expansion timescales during the explosion. Other nucleosynthetic signatures of
PISNe are $\mbox{[Mg,Si/Na,Al]}\gg 0$; $\mbox{[Si,S/C]}\sim 1$--$1.5$;
$\mbox{[Zn/Fe]}\ll 0$ \citep{heger:02,umeda:02,karlsson:11}.  Hence metal-poor
stars pre-enriched by PISNe can be identified by these chemical signatures.

In the Sloan Digital Sky Survey (SDSS), candidate metal-poor stars are
selected by means of $ugriz$ photometry; therefore, the above-mentioned bias
against PISNe pre-enriched stars is not present in the sample of stars for
which low-resolution (i.e., $R=\lambda/\Delta\lambda \sim 2000$) spectroscopy
was obtained.

The seventh data release of SDSS (DR7) contains $460,000$ stars; for about
half of them, low-resolution spectra have been acquired. According to
\citet{karlsson:08}, the number fraction of second-generation stars below
$\mbox{[Ca/H]}=-2.0$ with a dominant (i.e.$>90$\,\%) contribution from PISNe
is $~10^{-4}$ to $5 \times 10^{-4}$. About $10$\,\% of these PISN-dominated
stars are predicted to have $\mbox{[Fe/Ca]}\sim-1$. Therefore, according to
the current model \citep{karlsson:08}, at least two stars and up to 9--16
stars with dominant pre-enrichment by PISNe could be found in DR7.

After verifying our method by this pilot study, we will also apply the method
to DR8 and future SDSS data releases, as well as to the several million
stellar spectra to be obtained in the course of the LAMOST (The Large Sky Area
Multi-Object Fiber Spectroscopic Telescope) Galactic survey \citep{zhao:2012}, which will
provide more candidates. If we can find any stars pre-enriched by PISNe, this
would provide observational evidences for their existence; otherwise an upper
limit for their space density can be derived.

In this paper we report on the candidate selection applied to SDSS DR7. In
section \ref{sect:sample}, we give a brief description of the star sample from SDSS DR7;
Section \ref{sect:SSPP} introduces the metallicities from SSPP; the data analysis process
and candidates selection are described in section \ref{sect:metallicity} and \ref{sect:candidate}, respectively;
 Finally a conclusion is given in section \ref{sect:conclusions}.

\section{The SDSS Sample}\label{sect:sample}


The observations forming the basis of SDSS-DR7 were carried out during an
eight-year period with a dedicated 2.5-m telescope located at Apache Point
Observatory in Southern New Mexico. DR7 includes phtometric data of 460,000
stars and over 300,000 stellar spectra covering more than 8200 square degrees
of the sky. The wavelength coverage is $3800$--$9200$\,\AA, with a resolving
power of $R=1800$--$2200$ and $S/N>4$ per pixel at $g=20.2$. We extracted
224,080 stars from the table $star$ joining to $sppParams$ in SDSS-DR7 using
$CasJobs$, which is the online workbench for extracting large scientific
catalogs. The number reduced to 17,623 when narrowing the temperature range to
$4500$--$6500$\,K, and requiring the average signal-to-noise ratio ($S/N$) per pixel of more than 20
in the wavelength range $4000$--$8000$\,{\AA}. We adopted this temperature range by 
inspection of the feasibility of our method on different spectra. After the data analysis, about
5,000 spectra were rejected due to lack of flux points in the
wavelength range which contains the iron lines we use. The final sample to
which we applied our selection criteria hence consists of 12,304 stars.

%

When this work was under preparation, the SDSS Data Release 8 (DR8) was
published. Besides adding more spectra of stars to the data release, the final
adopted values of $T_{\mathrm{eff}}$ and $\log g$ are only slightly different
from DR7, while the [Fe/H] estimates are in general improved, especially at
the low-metallicity ($< -3.0$) and high-metallicity ($>0.0$) ends, due to a
re-calibration of the NGS1 and NGS2 parameter estimation approaches (see
Table~\ref{Tab:ssppmetallicity}), and changes in $S/N$ and
$(g-r)_{\mathrm{0}}$ \citep[]{smolinski:11}. Therefore, the primary sample
selection based on $T_{\mathrm{eff}}$ and $\log g$ is still valid, but we
consider the [Fe/H] from DR8 instead of DR7. The [Fe/H] from DR7 and DR8 are
both listed in Table \ref{tab:candidate}.  

\section{Metallicities derived by the SSPP}\label{sect:SSPP}

SSPP is the stellar parameter pipeline developed for the Sloan Extension for
Galactic Exploration and Understanding (SEGUE), which is a part of SDSS-II and
SDSS-III. This pipeline works on low-resolution spectroscopy and $ugriz$
photometry obtained during the course of SDSS-I, SDSS-II/SEGUE-1 and
SDSS-III/SEGUE-2, making use of multiple techniques to estimate fundamental
stellar atmospheres parameters (including effective temperature, surface
gravity, and metallicity), along with determining radial velocities for
AFGK-type stars \citep{lee:08a,lee:08b,allende:08,smolinski:11}.  Different
approaches were applied for each parameter, as summarized in
Table~\ref{Tab:ssppmetallicity}.  For [Fe/H], 12 different methods were
adopted in SSPP.


\begin{table}[t]
\centering
\caption{Metallicity estimation methods used by the SSPP}
\label{Tab:ssppmetallicity}
 \begin{tabular}{c c c}
 \hline\hline
 VALUE & METHOD & DESCRIPTION\\
 \hline
    feh1 & NGS2   & $\chi ^{2}$ minimization technique using Kurucz NGS $\alpha$-enhanced grid$^{1}$ \\
    feh2 & NGS1   & $\chi ^{2}$ minimization technique using Kurucz NGS non-$\alpha$-enhanced grid$^{1}$ \\
    feh3 & ANNSR  & Neural network trained on synthetic spectra$^{1}$ \\
    feh4 & ANNRR  & Neural network trained on real spectra$^{2}$ \\
    feh5 & CaIIK1 & Spectral fitting in the range $\lambda=3850$-$4250$\,{\AA}, using the NGS1 grid$^{1}$ \\
    feh6 & CaIIK2 & \ion{Ca}{II} K line strength combined with a broadband colour$^{3}$ \\
    feh7 & CaIIK3 & Neural network approach using the Ca~K line index K24 and $g-r^{1}$ \\
    feh8 & ACF    & Autocorrelation function proportional to the frequency and strength of weak metallic lines $^{3}$ \\
    feh9 & CaIIT  & Strength of the Ca II triplet lines and $B-V$; neural network approach$^{4}$ \\
    feh10 & WBG   & Equivalent width of the of \ion{Ca}{II} K line and comparison of weaker metallic lines$^{5}$ \\
    feh11 & k24   & matching of synthetic spectra using k24 grid of flux with the observed \\
          &       & flux in the region $\lambda=4400$--$5500$\,{\AA}, including $g-r$$^{6}$ \\
    feh12 & ki13  & matching synthetic spectra using ki13 grid of flux without colour with \\
          &       & the observd flux in the region $\lambda=4400$-$5500$\,{\AA}$^{1}$ \\
    feha & ADOP   & Adopted [Fe/H] value, combination of good estimates\\
 \hline
 \end{tabular}
\begin{list}{}{}
\item[$^{\mathrm{1}}$] \citealp{lee:08a} 
\item[$^{\mathrm{2}}$] \citealp{refiorentin:07}
\item[$^{\mathrm{3}}$] \citealp{beers:99}
\item[$^{\mathrm{4}}$] \citealp{cenarro:00a,cenarro:00b}
\item[$^{\mathrm{5}}$] \citealp{wilhelm:99}
\item[$^{\mathrm{6}}$] \citealp{allende:06}
\end{list}
\end{table}

Each of these methods only works for a certain colour range and $S/N$
level. More detailed information can be found in Table~5 of
\citet{lee:08a}. Among these approaches, $\emph{feh6}$ and $\emph{feh7}$ use
the strength of \ion{Ca}{II}~K line in combination with a broadband colour to
estimate [Fe/H] \citep{beers:99,lee:08a}. $\emph{feh9}$ was obtained by
measuring the integrated strength of \ion{Ca}{II} triple features, which are
also sensitive to $\log g$, along with the de-reddened $B-V$ colour
\citep{cenarro:00a,cenarro:00b}. $\emph{feh5}$ applied the NGS1 grid to a
short wavelength window covering \ion{Ca}{II}~K, Balmer lines, and a
\ion{Ca}{I} line, to estimate the metallicity, temperature and surface
gravity, respectively \citep{lee:08a}. $feh10$ uses a combination of the
equivalent width of \ion{Ca}{II}~K line and a comparison of weaker metallic
lines to synthetic spectra \citep{wilhelm:99}. The adopted $feha$ containing
the information of Ca abundance as well as other weak metal lines, is much
closer to the theoretical definition of metallicity.  A flag value
0 is assigned to the result from each method if the star does not satisfy the
range of validity for this method, or 1 if it is defined as an outlier after
matching with the synthetic spectra \citep[See][]{lee:08a,smolinski:11}.

In order to make the comparison more sensitive to the deviation between the
abundance of calcium and iron, we use the averaged value of $feh6$ and $feh7$,
if they are both valid, as indicated by a flag value of 2; or the valid one 
of them. Otherwise, one of $feh9$, $feh10$ and $feh5$ would be used, depending on 
the given priority, which is based on the correlation degree to Ca abundance. In
the case that none of them passed the synthetic spectra matching check, the
adopted $feha$ was used.

\begin{figure}[htp]
\resizebox{\hsize}{!}{\includegraphics[bb = 90 230 480 730, clip = true]{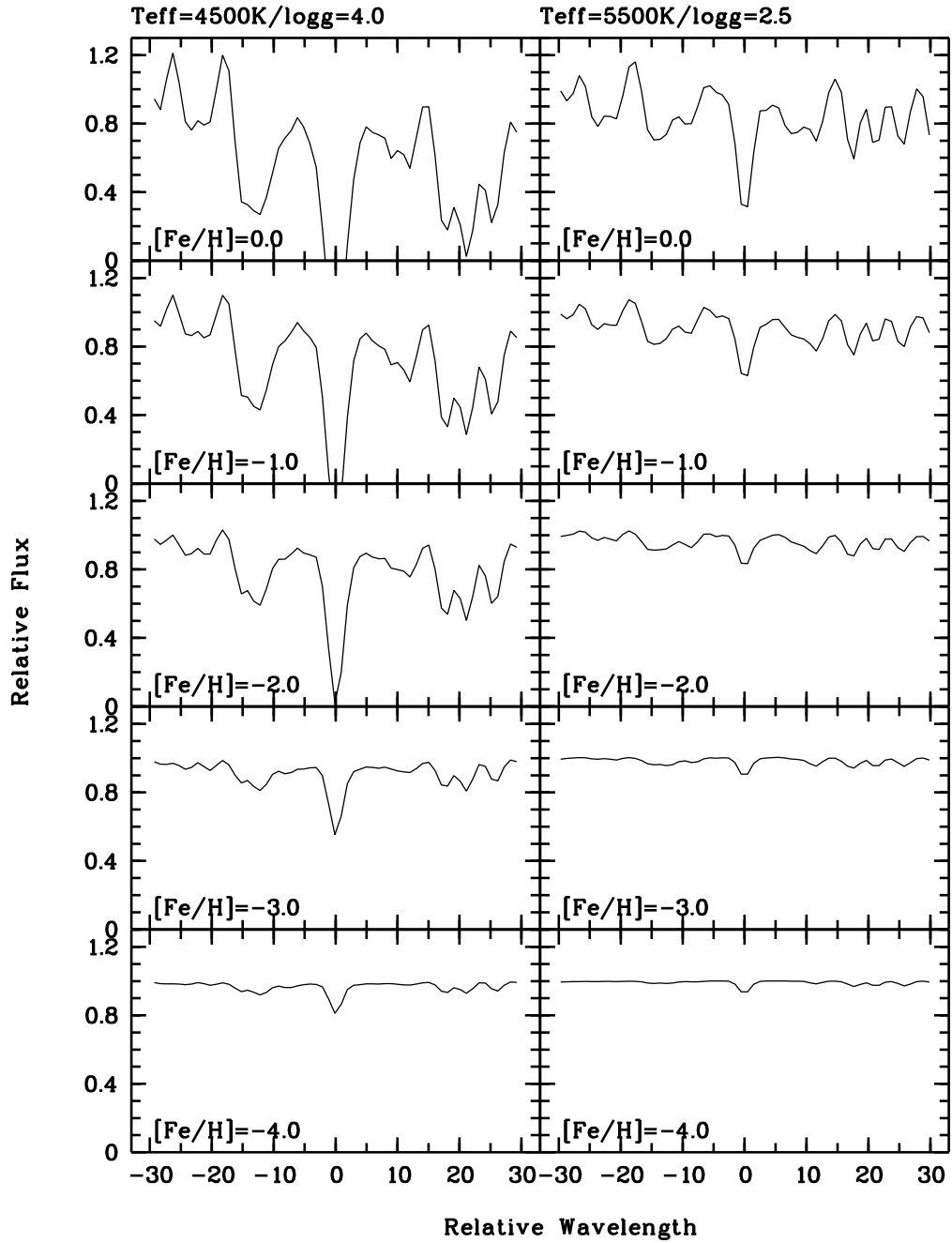}}
\caption{The result of coadding iron lines in synthetic spectra with different
  [Fe/H] for $T_{\mathrm{eff}}=4500$/$\log g=4.0$ and
  $T_{\mathrm{eff}}=5500$/$\log g=2.5$, respectively.  The line center is
  shifted to a wavevelength of zero in the rest frame.}
\label{fig:coadd}
\end{figure}

\section{Metallicity from coadded iron lines}\label{sect:metallicity}

The metallicity $Z$ of a star refers to the proportion of its matter made up of chemical 
elements heavier than helium. Iron is often used as the reference element for
metallicity, since it is the end product of exothermic nucleosynthesis in
stars, and because numerous iron lines are present in the optical wavelength
region, so that its abundance can easily be measured in high-resolution
spectra of cool stars.  However, if the metallicity is to be determined from
low resolution spectra, one has to resort to the much stronger \ion{Ca}{II}~H
and K lines and assumptions on the relation between [Fe/H] and [Ca/H]
\citep[See, e.g.,][]{beers:99}.

\subsection{Selecting iron lines}\label{subsect:check}


\begin{table}[t]
\caption{The two Fe I lines we $\mathrm{coadded}^\mathrm{a}$}
\label{Tab:ironline}
\centering
 \begin{tabular}{c c c c c}
 \hline\hline
 $\lambda$ [{\AA}] & Species & $W_{\sun}$ [{\AA}] & $\log gf$ & $\xi$ [eV]\\
 \hline
    $4045.82$ & Fe I & 1.17 & 0.280 & 1.485 \\
    $4383.56$ & Fe I & 1.01 & 0.200 & 1.485 \\
 \hline
 \end{tabular}
\begin{list}{}{}
\item[$^{\mathrm{a}}$]log($gf$) and lower excitation 
energy are from Vienna Atomic Line Database (VALD). The equivalent widths (EW) 
are the line strength in the Sun. \\
 http://www.astro.univie.ac.at/${ }^{\sim}\mathrm{vald}$/
\end{list}
\end{table}

We searched for strong iron lines detectable in SDSS spectra, using
identifications of lines in the Solar spectrum \citep{moore:66}.  Because of
the limited quality of the SDSS spectra in terms of resolving power and $S/N$,
most of the Fe lines that are very strong in the Sun are difficult to detect
in the SDSS spectra.  After attempts to identify these lines in spectra of
stars with different $T_{\mathrm{eff}}$, $\log g$ and [Fe/H], the lines with
equivalent widths below $1$\,{\AA} were rejected. Additionally, because some
SDSS spectra have no flux data around $3820$\,{\AA}, and also because lines at
the extreme blue end with wavelength smaller than $4000$\,{\AA} are usually
blended with other lines or completely hidden in the noise, finally, two
relatively strong iron lines are left: Fe~I~$4045.82$\,{\AA}, and
Fe~I~$4383.56$\,{\AA}. Their atomic data are listed in
Table~\ref{Tab:ironline}.

Fig.~\ref{fig:coadd} displays the coadded iron lines in synthetic spectra of
stars with different metallicities, which demonstrates that they are strong
enough to be detected at {[Fe/H]}$\ge-4.0$ and $T_\mathrm{eff}$~$\le6500$K. We
hence restricted the application of our method to stars in this
$T_\mathrm{eff}$ range. Note that additional information on the metallicity is
contained in the continuum around the coadded Fe~I lines. Our method
implicitly makes use of this information via the continuum bands that are used
to scale the synthetic spectra to the observed ones.

\subsection{Normalization}\label{subsect:norm}

Continuum normalization of a spectrum is always a delicate task, even more so
when the spectra to be normalized only have a low resolution, and when the
task must be performed automatically for stars with a variable range of
effective temperatures, which is the case in our application. Because the two
iron lines we use both lie in the wavelength range of $4000$--$4500$\,{\AA},
only a short wavelength range in the spectrum covering the two iron lines is
used.  This makes the normalization easier, since a smaller wavelength region
has to be modeled. We normalize the spectrum in the region
$\lambda=3820$--$5800$\,{\AA}, which is a bit larger than the range we use for
following anaysis, because we want to include more continuum points in the
spectrum for a more accurate continuum determination. We adopt the same
division of the spectrum as that used for two seperate continuum fits in SSPP 
(Lee et al. 2008), because we want to be consistent with the
SSPP analysis process as much as possible, which will make the comparison of
the results more robust.

\begin{figure}[htp]
\resizebox{\hsize}{!}{\includegraphics[bb = 80 310 510 740, clip = true]{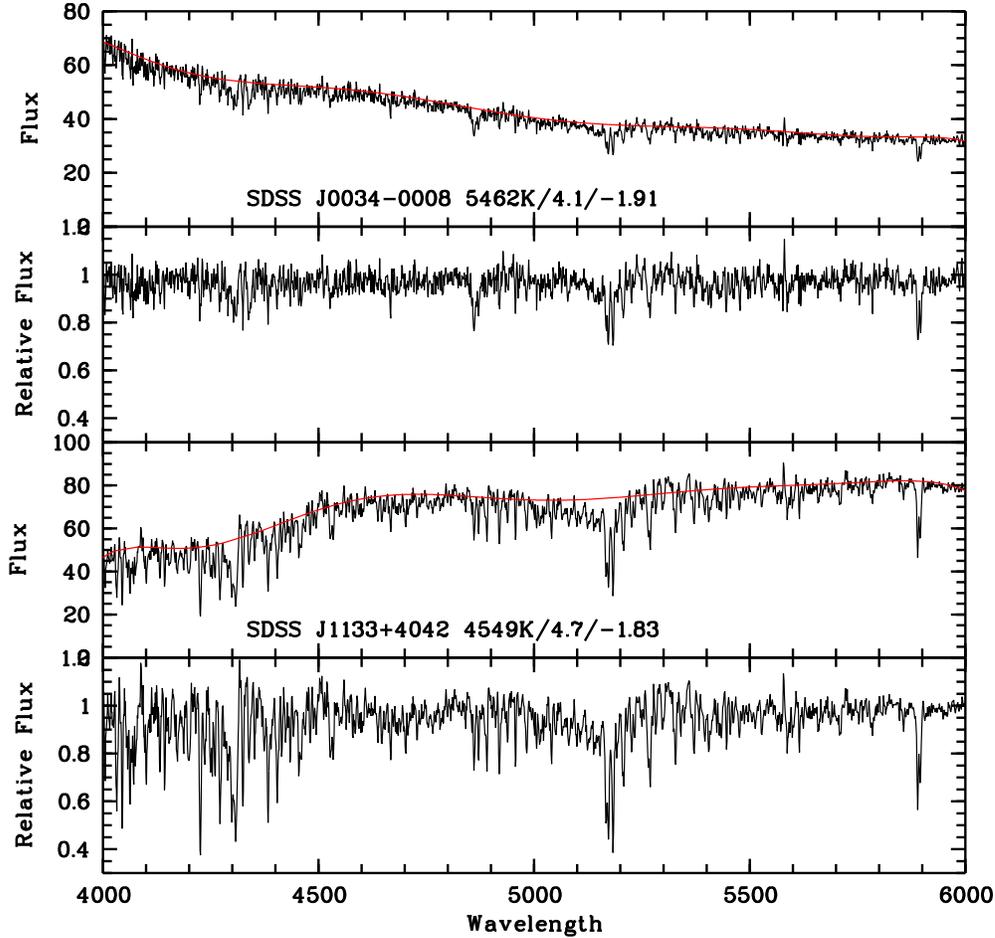}}
\caption{Continuum determination for SDSS spectra of our two candidates: SDSS J0034$-$0008 and SDSS J1133$+$4042.}   
\label{fig:norm}
\end{figure}

In order to determine an approriate continuum fit, some strong lines are
removed from the spectrum, i.e., Balmer lines, Ca II K lines, Ca I line, Mg Ib
lines, G-band and iron lines, after which the spectrum is iteratively fitted
to a ninth-order polynomial. We then rejected the points outside the $3\sigma$
range of the fitted function after the first fit, which can also remove the
cosmic rays.  In the following iterations, considering the numerous weak lines
left in the spectrum, only points within $2\sigma \sim -0.5\sigma$ range of
the fitted function were kept. The final continuum was obtained after four
interations. Fig.~\ref{fig:norm} shows two examples of the normalization for
 two observed SDSS spectra, among our candidates.

The synthetic spectra we use were kindly provided by Y. S. Lee (priv. comm.).
The wavelength coverage is $3000$--$10,000$\,{\AA} in 0.01\,{\AA} steps, and
they were generated using Kurucz models and the spectrum synthesis program
TURBOSPECTRUM. The spectra cover a wide range of the stellar parameters
T$_{\mathrm{eff}}$, $\log g$ and [Fe/H], from $4000$\,K to $9750$\,K, $0.0$ to
$5.0$ dex and $-4.0$ to $0.5$\,dex, in steps of $250$\,K, $0.5$\,dex, and
$0.5$\,dex, respectively. The synthetic spectra are processed in exactly the
same way as the SDSS spectra, in order to reduce systematic errors.

\subsection{Coadding iron lines}\label{subsect:coadding}

\begin{figure}[hp]
\resizebox{\hsize}{!}{\includegraphics[bb =20 50 550 300, clip = true]{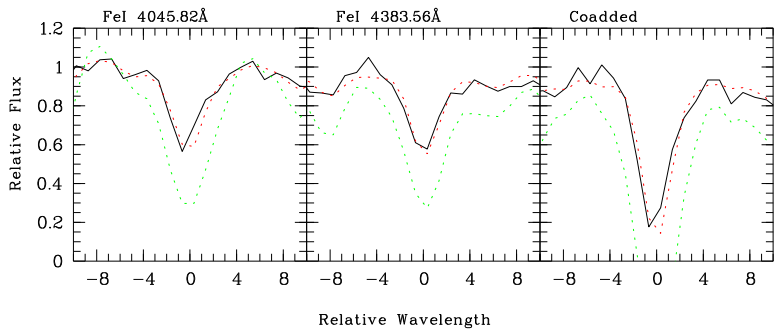}}
\caption{Line strength before and after coadding two iron lines of an observed
  spectrum for one of our candidates (solid, SDSS J2118$-$0613) and the synthetic spectra (dashed,
  $T_{\mathrm{eff}}=4750$\,K/$\log g=4.5$/$\mbox{[Fe/H]}=-2.0/0.0$ dex). The red and green dashed line correspond 
to the best fit synthetic spectrum from our analysis ([Fe/H]=$-2.04$) and from SSPP ([Fe/H]=$-0.10$), respectively.}
\label{fig:comcoadd}
\end{figure}

Because the wavelength scale used in SDSS spectra is based on vacuum, the
first step before coadding the iron lines is to transform the wavelength scale
of the original SDSS spectrum to an air-based wavelength scale, and after
shifting the spectrum to a zero-velocity rest frame using the radial velocity
estimated from matching ELODIE template spectra \citep{prugniel:01,moultaka:04,adelman-mccarthy:08}, 
the two iron lines are shifted to a wavelength of zero. The radial velocity from ELODIE is
reported to be the best available estimate by previous experiences
\citep{lee:08a}. Linear interpolation is used to put the spectral regions of the
two lines on the same pixel scale, before the spectra are coadded. To speed up
the analysis process, only ranges of $60$\,{\AA} centered on the two iron
lines were used.  Fig.~\ref{fig:comcoadd} shows the iron lines detected in the
spectra and the strength of the lines after coadding, for both observed and
synthetic spectra. The synthetic spectra corresponding to the best-fit metallicity from 
our analysis and from SSPP are also plotted for comparison. It is clearly seen that the 
adopted SSPP metallicity is too high for iron lines. It is likely that the SSPP 
result is dominated by other chemical features other than iron, e.g., \ion{Ca}{II}, 
\ion{Mg}{II}, or even the molecular carbon bands for some cases, which we will discuss more later.

\subsection{Metallicity determination}\label{subsect:chisquare}

Based on the T$_{\mathrm{eff}}$ and $\log g$ from SSPP, the corresponding
synthetic spectrum was chosen, and then it was degraded and interpolated to
the same resolving power ($R \sim 2000$) and wavelength scale as SDSS
spectrum.  Since the resolution does not change much in the wavelength range
we use in this work, we adopted a constant wavelength step width. Then we used
a reduced $\chi^2$ criterion to get the best-fit metallicity by minimizing the
difference between the observed and synthetic coadded flux. We used a range of
$10$\,{\AA} of the coadded spectrum, which only covers the coadding feature,
centered on the line, in order to reduce the impact from other lines and noise
in the rest part of the spectra. A third-order polynomial is adopted to fit
all the $\chi^2$ values from different metallicities, and the metallicity
coresponding to the minimum $\chi^2$ is taken as the best-fit value.

\subsection{Error estimation}

In order to estimate the uncertainty of our method, we analyzed 1069 stars belonging to different 
globular clusters (GCs) and open clusters (OCs). The spectroscopic data were obtained during SEGUE observations and they
 were used to check the validation of the SEGUE target selection and SSPP \citep{lee:08b,smolinski:11}.
Only GC members with 4500\,K$\leq T_\mathrm{{eff}}\leq$6500\,K and $S/N>20$ were selected in 
consistency with our sample, as described in Section \ref{sect:sample}. Finally, 724 stars 
were considered for eight GCs (M3, NGC 5053, M53, M71, M92, M15, M2, and M13) and five OCs 
(NGC 6791, M35, M67, NGC 2420, and NGC 2158). The high-resolution metallicity results are from the 
work of \citet{harris:96,dias:02} and \citet{boesgaard:09}. By comparing the 
metallicities from our work with the averaged [Fe/H] from high-resolution analysis for each cluster, 
we estimated the uncertainty of our method to be $\sim 0.40$ dex.

\section{Candidates selection}\label{sect:candidate} 

\begin{figure}[h]
  \centering
   \includegraphics[bb = 60 250 600 500, clip = true]{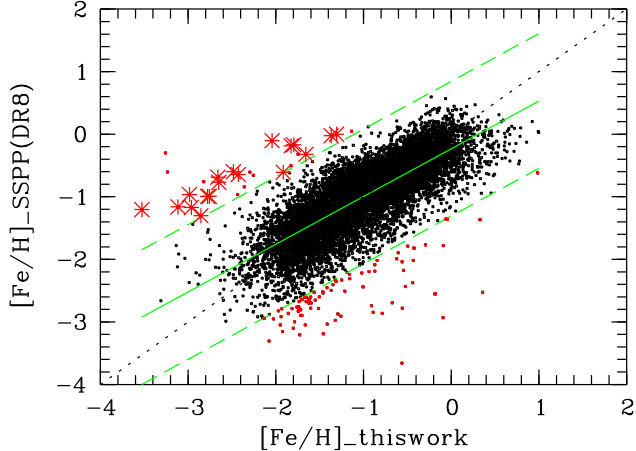}
  \caption{{\small Comparison between metallicities from SSPP and our method.
      Those $3\sigma$ outliers with S/N $>$ 25 (red stars) on the upper side show a possible
      Ca enhancement. Both the upper and lower $3\sigma$ outliers are highlighted in red dots.}}
\label{fig:resultcom}
\end{figure}

Due to the differences in the methods used, together with the weakness of the
Fe~I lines, the limited quality of the SDSS spectra, and uncertainties incured
by the automated analysis, our [Fe/H]$\_{\chi ^2}$ estimated directly from
Fe~I lines expectedly deviates from the SSPP metallicities
[Fe/H]$\_{\mathrm{sspp}}$. The average deviation is $0.29\pm 0.25$\,dex. 
As shown in Fig. \ref{fig:resultcom}, at the high and low metallicity ends, larger scatters 
appear in the comparison. A linear fit 
[Fe/H]$_{\mathrm{SSPP}}$ = $0.76 \times$ [Fe/H]$_{\mathrm{this work}}-0.23$ is performed,  
 with a standard deviation of 0.35 dex, which demostrates that our result is systematically higher at [Fe/H]$>-1.0$. 
 The $3\sigma$ outliers based on the 
linear fit are interesting, despite possible origination from
errors. In this work, we care about those objects falling outside $3\sigma$
with likely Ca-enhancement, which might be potential candidates for PISNe
enriched metal-poor stars. By means of visual inspection of the original and
coadded spectra, we rejected false positives from the candidate sample, i.e.,
spectra suffering from overestimated signal-to-noise ratios or poor continuum
determination. Finally, 18 stars were selected as Ca-enhanced candidates. They are listed in
Table~\ref{tab:candidate}, and shown as red stars in
Fig.~\ref{fig:resultcom}.  

The spectra of outliers with lower SSPP metallicity were also inspected, and we found that 
the large departure is mainly due to low S/N and hence poor continuum determination. 


In addition, the outliers are mostly cooler
($T_{\mathrm{eff}}<5000$\,K) and metal-poor stars ([Fe/H]$\le-1.00$), therefore, according to 
\citet{lee:08a} the estimated metallcity from Ca lines in this case might be underestimated by $\sim 0.5$ dex,
indicating the true departure from our derived results would be even larger.

Although We note that a large fraction of the used [Fe/H] values from the SSPP
are the final adopted value $feha$, which contains the information of Ca and
other metallic lines, e.g., \ion{Mg}{I}, \ion{Fe}{I}, \ion{Na}{I}, the true
difference (excluding uncertainty) between the two metallcities can still
reflect the deviation of abundance between Fe lines and other metal lines,
especially Ca lines, since some weak Ca lines might be used and the final
valid metallicity estimators were selected after matching with the synthetic
spectrum in the wavelength range of $3850$--$4250$\,{\AA} and
$4500$--$5500$\,{\AA} \citep{smolinski:11}.

\begin{figure}[htp]
\resizebox{\hsize}{!}{\includegraphics[bb = 20 580 600 790, clip]{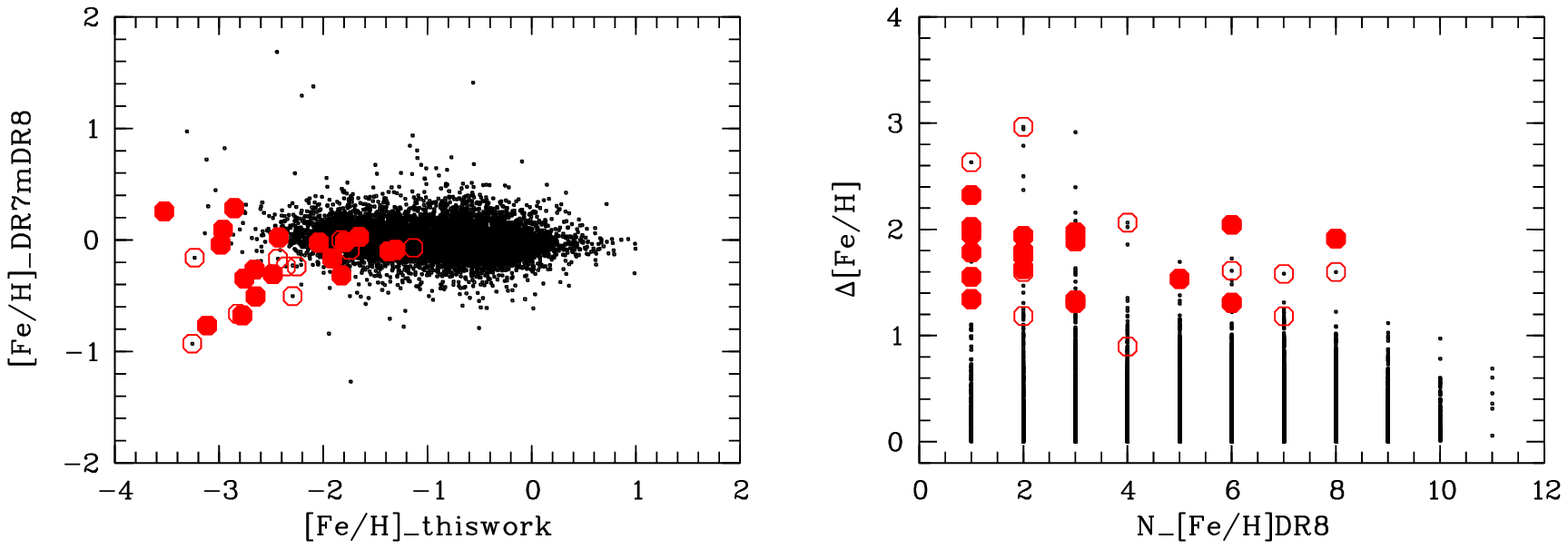}}
\caption{Left: Metallicity deviation of DR7 from DR8, distributing with our
  derived metallicity. Right: Metallicity deviation of DR8 from our method, distributing
  with the number used to give the final metallicity in SSPP DR8. Black dots:
  the whole sample; red circles: $3\sigma$ outliers with higher Ca; red filled
  circles: selected candidates.}
\label{fig:distributea}
\end{figure}

We inspected \ion{Ca}{I} 4226\,\AA~,  \ion{Ca}{II} H and K lines 
in the spectra of our candidates, and no apparent 
enhancement of Ca abundance was found on \ion{Ca}{I} 4226\,\AA~ line, while 
for  \ion{Ca}{II} H and K lines, the situation is complicated due to the saturation and 
blendings. Furtherly, we are not clear about 
the sensitivity of Ca abundance on the calcium line strength, and whether 
these stars are influenced by chromospheric activities, thus better quality data 
are needed to measure the accurate abundances and look into these questions.
 In Fig. \ref{fig:distributea}, our metallicity
results also indicate that there are larger differences bewteen the [Fe/H]
values of DR7 and DR8 at the low metallicity tail compared to the differences
at higher metallicity. The number of stars with a large offset in the
comparison between DR8 and our derived [Fe/H] increases when a smaller number
of estimators is used, and fewer estimators ($\leq3$) are used at lower metallicity ([Fe/H]$<-3.0$).

%
 
\setlength{\tabcolsep}{1mm}
\begin{table}
\begin{center}
\caption[]{Candidates of Metal-poor Stars Pre-enriched by PISNe.}\label{tab:candidate}


 \begin{tabular}{lclclclrrrr}
  \hline\noalign{\smallskip}
 Star & Plate & MJD  &  Fiberid &  S/N  &  $T_{\mathrm{eff}}$  & $\log g$ & [Fe/H] & [Fe/H]$^{a}$ & [Fe/H] & Note $^{b}$\\
      &       &      &          &       &   (K)                &          &    DR7 & DR8(Ca) & this work & \\
  \hline\noalign{\smallskip}
 SDSS J081157.13+143533.0 & 2270 & 53714 & 253  & 52 & 4750 & 4.0 & $-0.95$ & $-1.20$ & $-3.52$  & adop, 1\\ 
 SDSS J015315.91-084216.8 & 0665 & 52168 & 489  & 35 & 4822 & 3.6 & $-1.00$ & $-0.96$ & $-2.98$  & adop, 1 \\
 SDSS J211851.04-061319.6 & 0639 & 52146 & 538  & 47 & 4795 & 4.4 & $-0.13$ & $-0.10$ & $-2.04$  & adop, 2 \\
 SDSS J030209.90-062704.0 & 0458 & 51929 & 365  & 33 & 4657 & 4.2 & $-0.62$ & $-0.65$ & $-2.43$  & adop, 1 \\
 SDSS J121144.01+011951.3 & 0517 & 52024 & 003  & 65 & 4798 & 4.5 & $-0.19$ & $-0.17$ & $-1.79$  & adop, 2 \\
 SDSS J205923.08-065208.2 & 0636 & 52176 & 064  & 42 & 5298 & 4.0 & $-0.91$ & $-0.60$ & $-2.48$  & wbg, 3 \\
 SDSS J080805.07+183352.0 & 1923 & 53319 & 595  & 31 & 4830 & 4.8 & $-1.19$ & $-0.99$ & $-2.76$  & caiik2, 8\\
 SDSS J112238.18+434555.3 & 1365 & 53062 & 144  & 26 & 4808 & 4.5 & $-1.11$ & $-0.77$ & $-2.65$  & caiit, 6\\
 SDSS J164435.94+203659.1 & 1569 & 53168 & 232  & 36 & 4665 & 4.6 & $-0.32$ & $-0.32$ & $-1.66$  & caiit, 6\\
 SDSS J075800.22+435257.9 & 0437 & 51869 & 374  & 60 & 4588 & 4.9 & $-0.13$ & $-0.02$ & $-1.37$  & adop, 1 \\
 SDSS J111218.53+262543.8 & 2212 & 53789 & 636  & 43 & 4624 & 5.0 & $-0.09$ & $0.00$ & $-1.31$   & adop, 3 \\
 SDSS J113344.77+404238.4 & 1443 & 53055 & 090  & 36 & 4549 & 4.7 & $-0.61$ & $-0.19$ & $-1.83$  & caiik1, 5 \\
 SDSS J155219.79-001022.7 & 0342 & 51691 & 038  & 35 & 4662 & 3.9 & $-1.70$ & $-0.99$ & $-2.78$  & caiik3, 2 \\
 SDSS J104459.32+360554.7 & 2090 & 53463 & 273  & 33 & 5877 & 4.3 & $-0.95$ & $-0.69$ & $-2.66$ & adop, 3 \\
 SDSS J095423.10+240410.8 & 2341 & 53738 & 019  & 27 & 5004 & 3.6 & $-1.93$ & $-1.16$ & $-3.12$ & adop, 1 \\
 SDSS J230130.60+001920.9 & 0380 & 51792 & 499  & 30 & 4798 & 4.5 & $-1.08$ & $-1.17$ & $-2.96$ & adop, 2 \\
 SDSS J153642.56+210747.4 & 2166 & 54232 & 266  & 31 & 5489 & 4.2 & $-1.02$ & $-1.30$ & $-2.85$ & adop, 1 \\
 SDSS J003416.18-000849.9 & 0392 & 51793 & 224  & 27 & 5462 & 4.1 & $-0.75$ & $-0.61$ & $-1.91$  & wbg, 3 \\

\noalign{\smallskip}\hline
\end{tabular}
\begin{list}{}{}
\item[$a:$]~{[Fe/H] of DR8(Ca) refers to the Ca-related metallicity from SSPP(DR8) used in this work.}
\item[$b:$]~{This column list the method used to determine [Fe/H]-DR8(Ca) and the number of valid methods in SSPP.}
\end{list}
\end{center}
\end{table}

In general, our method to derive [Fe/H] is reliable, as seen from their good
correlation with the metallicity of different approaches used in SDSS-DR8.  We
also note that only a few selected outliers have valid result from methods
using \ion{Ca}{II} lines, which weaken the correlation with Ca-enhancement.
We do not exclude the possibility that the large departure of our [Fe/H] from
those derived with the SSPP is caused by errors in either or both of the two
estimates. Therefore, it is necessary to acquire spectra of higher resolution
and $S/N$ to verify our selection method, and for possibly identifying via
their abundance patterns stars pre-enriched by PISNe. To this end,
medium-resolution ($R\sim6000$--$11,000$) of our candidates are currently
being obtained with the XSHOOTER spectrograph attached to the Unit Telescope 2
of the Very Large Telescope of the European Southern Observatory.

\section{Conclusions}\label{sect:conclusions}

Very massive ($140\,M_\mathrm{\sun} \leq M \leq 260\,M_\mathrm{\sun}$), first
generation stars end their lifes as PISNe, which have been predicted by
theories, but no strong observational evidence for their existence in the
early Universe has yet been obtained. The theoretically predicted chemical
fingerprints of PISNe are high overabundances of Ca with respect to Fe and the
ratio of these elements in the Sun, and a strong odd-even effect. We carried
out a pilot study on identifying metal-poor stars with strong Ca-enhancement,
by comparing the metallicities determined with the SDSS SSPP containing
information on the Ca abundance (i.e., those estimators which involve the
\ion{Ca}{II}~H and K lines or the \ion{Ca}{II} infrared triplet) with
metallicities we derived directly from iron lines.  We used a coadding
technique to increase the $S/N$ of the SDSS spectra. A $\chi ^{2}$
minimization method was applied to search the best matched metallicity from a
series of synthetic spectra with known atmosphere parameters.  Our derived
metallicities correlate well with those of the SSPP, with a linear fit
[Fe/H]$_{\mathrm{SSPP}}$ = $0.76 \times$ [Fe/H]$_{\mathrm{this work}}-0.23$. By applying a
$3\sigma$ criterion and visual inspection, 18 candidates were selected. Once
the nature of these candidates is determined and our selection method is
verified, we will apply our method to the low-resolution stellar spectra
obtained with the LAMOST telescope.

\begin{acknowledgements}
We give thanks to Young Sun Lee for kindly offering us the synthetic spectra; 
Torgny Karlsson and Haining Li are acknowledged for their helpful comments on the 
manuscripts; we appreciate the stimulating discussions with Yuqin Chen and her helpful 
suggestions and comments on the manuscripts. 
J.R. and N.C. acknowledge financial support by the Global Networks program of
Universit\"at Heidelberg, and by Deutsche Forschungsgemeinschaft through grant
CH~214/5-1 as well as Sonderforschungsbereich SFB 881 ``The Milky Way System''
(subproject A4). J.R. and G.Z. acknowledge the support by NSFC grant No. 10821061.
J.R. acknowledges partial financial support from the Shandong University Fund
for Graduate Study Abroad. 
\end{acknowledgements}

\bibliographystyle{aa}
\bibliography{cosmology,sdss}
\label{lastpage}

\end{document}